\documentclass[twocolumn, amsmath, trackchanges, twocolappendix]{aastex702}

\makeatletter
\providecommand{\@journalinfo}{}
\makeatother

\begin{document}

\title{\vspace*{-45pt} \large Fast and Differentiable MHD Surrogate for Solar Wind using Neural Operator}

\author[orcid=0000-0001-8265-6254,sname='Mayank']{Prateek Mayank}
\affiliation{University Corporation for Atmospheric Research, CPAESS, Boulder, CO 80301, USA}
\affiliation{Space Weather TREC, University of Colorado, Boulder, CO 80303, USA}
\email[show]{pmayank@ucar.edu}  

\author[orcid=0000-0002-7862-6383, sname='Camporeale']{Enrico Camporeale} 
\affiliation{Queen Mary University of London, London E1 4NS, United Kingdom}
\affiliation{Space Weather TREC, University of Colorado, Boulder, CO 80303, USA}
\email{} 

\author[orcid=0000-0002-4989-475X, sname='']{Thomas E. Berger} 
\affiliation{National Center for Atmospheric Research, High Altitude Observatory, Boulder, CO, 80301, USA}
\affiliation{Space Weather TREC, University of Colorado, Boulder, CO 80303, USA}
\email{} 

\author[orcid=0000-0003-1674-0647]{Zhenguang Huang}
\affiliation{University of Michigan, Ann Arbor, MI 48109, USA}
\email{} 

\author[orcid=0000-0001-8459-2100]{Gabor Toth}
\affiliation{University of Michigan, Ann Arbor, MI 48109, USA}
\email{}

\begin{abstract}
Global magnetohydrodynamic (MHD) simulations provide the complete plasma state of the heliosphere, but their computational cost limits applications requiring repeated model evaluations. We introduce SNOW (Solar Neural Operator for Wind), a fast and differentiable surrogate for global MHD solar wind. SNOW combines a rotational ballistic approximation, a spherical harmonic neural operator, and causal radial context to autoregressively propagate all eight MHD variables from \(20\) to \(240\,R_\odot\) through consecutive shell to shell transitions. Trained and evaluated using AWSoM simulations spanning a solar cycle, SNOW remains stable across the complete rollout. The model successfully reproduces the large scale plasma and magnetic field structures, including fast and slow wind streams, magnetic polarity sectors, vector orientations, fieldline connectivity, and bulk-flow structures. Additionally, the magnetic divergence and induction diagnostics remain comparable to the reference MHD solutions. The complete three-dimensional solution is generated in less than one second on a single GPU.
These results demonstrate the feasibility of fast neural emulation of global MHD solar wind solutions while retaining physically meaningful plasma and magnetic field structure.

\end{abstract}

%% Keywords should appear after the \end{abstract} command. 
%% The AAS Journals now uses Unified Astronomy Thesaurus (UAT) concepts:
%% https://astrothesaurus.org
%% You will be asked to selected these concepts during the submission process
%% but this old "keyword" functionality is maintained in case authors want
%% to include these concepts in their preprints.
%%
%% You can use the \uat command to link your UAT concepts back its source.
%% From the front matter, we move on to the body of the paper.
%% Sections are demarcated by \section and \subsection, respectively.
%% Observe the use of the LaTeX \label
%% command after the \subsection to give a symbolic KEY to the
%% subsection for cross-referencing in a \ref command.
%% You can use LaTeX's \ref and \label commands to keep track of
%% cross-references to sections, equations, tables, and figures.
%% That way, if you change the order of any elements, LaTeX will
%% automatically renumber them.

\section{Introduction} \label{sec:intro}
The solar wind is a continuous outflow of magnetized plasma from the Sun that fills interplanetary space and establishes the background state of the heliosphere. It provides the ambient medium through which solar transients propagate and sets the upstream plasma and magnetic field conditions that govern their interaction with planetary environments. Accurate knowledge of its kinematic, thermodynamic, and magnetic properties is therefore fundamental to heliospheric modeling and space weather forecasting.

Global three-dimensional (3D) magnetohydrodynamic (MHD) models have become the primary physics-based approach for investigating the large scale structure and evolution of the solar wind. By solving the governing MHD equations, these models describe the nonlinear interactions responsible for the formation and evolution of heliospheric structures. Multiple frameworks such as MAS \citep{mikic_MAS_1999, riley_MAS_2001, lionello_MAS_2023}, ENLIL \citep{dodstrcil_2003_modeling}, AWSoM \citep{van_AWSoM_2010, van_AWSoM_2014, Meng_AWSoM_2015}, and SWASTi \citep{mayank_2022_swastisw, prateekmayank_2023_swasticme, Mayank_2025_swasti} have consequently become important tools for both fundamental research and forecasting. Their ability to evolve the complete 3D kinematic, thermodynamic, and magnetic field state makes global MHD particularly valuable for providing boundary conditions to subsequent geospace models.

Despite their success, global MHD models remain computationally demanding. While the cost of an individual simulation can be manageable, many emerging applications require repeated model evaluations. Ensemble forecasting, uncertainty quantification, parameter estimation, and systematic sensitivity studies therefore become computationally expensive under conventional MHD workflows. Moreover, traditional numerical solvers are not naturally end-to-end differentiable, making gradient-based data assimilation, calibration against observations, and inverse optimization difficult to perform. These capabilities are increasingly important for advancing from deterministic simulations toward probabilistic and observation-constrained forecasting. This creates a growing mismatch between the capabilities required by the community and those that can be practically achieved with current numerical workflows.

Several computationally efficient alternatives have been developed to address parts of this problem. Reduced-physics propagation schemes \citep[HUX; ][]{riley_2011_hux}, semi-empirical models \citep[WSA; ][]{arge_2003_improved}, and more recently data-driven approaches \citep[WSA+; ][]{mayank2025neural} can provide rapid estimates of specific solar wind quantities, particularly the radial velocity. However, none of them evolve the complete 3D kinematic, thermodynamic, and magnetic field state. These quantities are themselves essential for space weather prediction and for self-consistent coupling to downstream models of the magnetosphere.

Recent developments in scientific machine learning offer a different route: rather than reducing the underlying physical system to a small set of variables, surrogate models can be trained to approximate the evolution produced by the numerical solver itself. In particular, neural operators learn mappings between functions and have emerged as a promising framework for approximating solution operators of partial differential equations (PDEs) \citep{FNO2020, DeepONet2021, NO_2023}. Their application to fluid dynamics has demonstrated that complex nonlinear evolution can be reproduced at a fraction of the computational cost of conventional numerical integration, including problems involving turbulence, transport, and other PDE systems \citep{Azizzadenesheli_2024, Burles_Camporeale_2026}. These developments raise the possibility of retaining the rich state information of physics-based simulations while substantially reducing the cost of repeatedly evolving that state.

Initial studies have begun extending this idea toward hydrodynamic (HD) and MHD systems. However, most existing demonstrations have focused on reduced-dimensional or idealized benchmark systems, restricted geometries, short temporal evolution, or subsets of the underlying physical variables \citep{degen2023perspectives, zhao2024, Li_2025, Mansouri2025}. Such capability has not yet been demonstrated for global 3D MHD systems.

In this work, we introduce SNOW (Solar Neural Operator for Wind), a fast and differentiable surrogate for 3D MHD solar wind in the inner heliosphere. Starting from the complete plasma state at $20R_\odot$, SNOW autoregressively propagates all eight MHD variables through consecutive spherical shells to $240R_\odot$. We demonstrate the capability of the model in physical fidelity of 3D reconstruction, vector orientation, and operational context. The complete heliospheric solution is generated in less than a second on a single GPU, compared with $>$100 CPU hours for the corresponding traditional global MHD simulation.

The rest of the paper is structured as follows. Section 2 describes the data representation, SNOW architecture, and autoregressive training strategy. Section 3 evaluates the reconstructed MHD solutions, and Section 4 discusses their physical fidelity, operational context, and current limitations and future extensions. The conclusion is in Section 5.

\section{Methodology} \label{sec:method}

\begin{figure*}
    \centering
    \includegraphics[width=\textwidth]{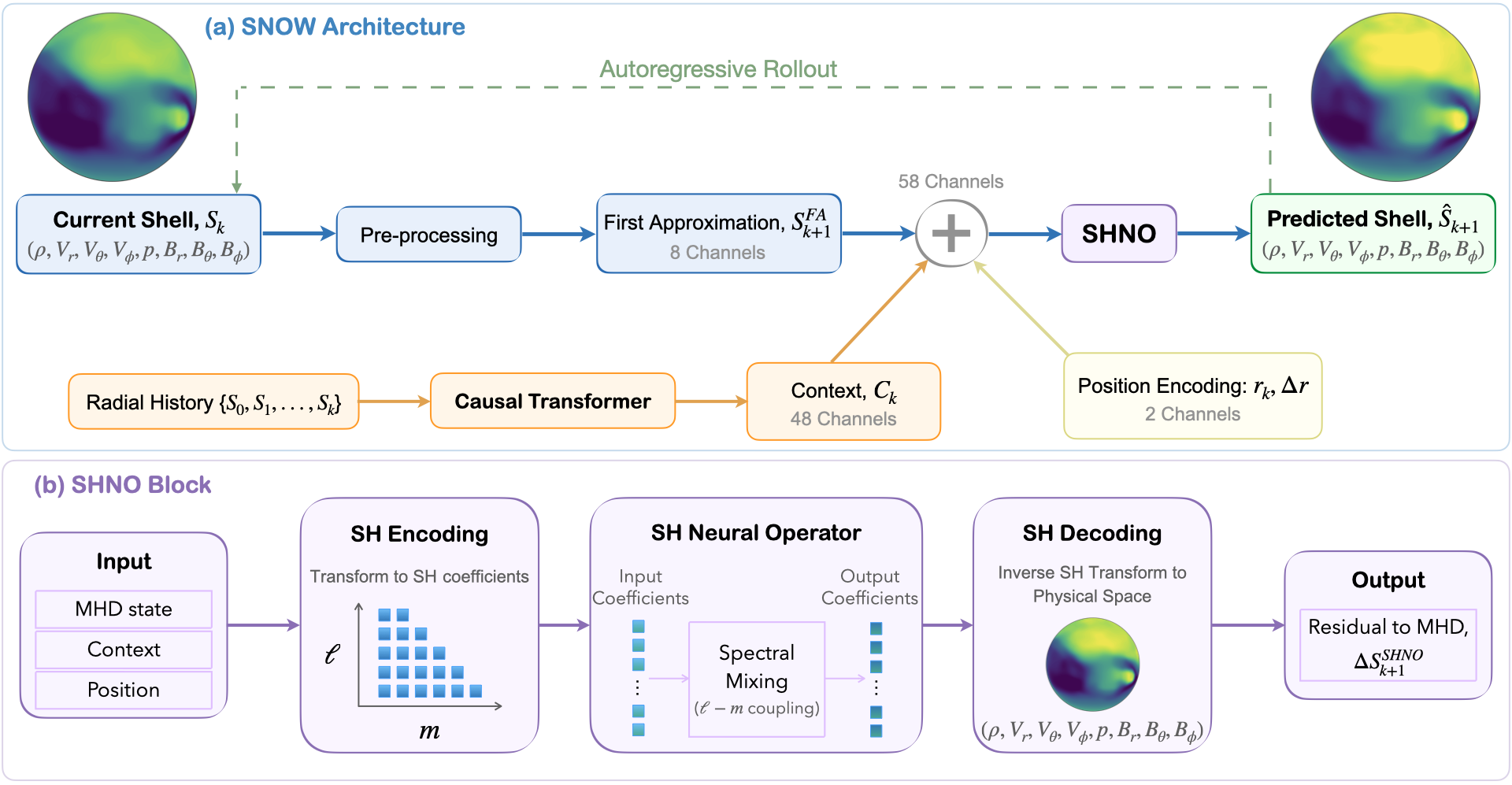}
    \caption{Architecture of SNOW for autoregressive radial propagation of the 3D MHD solar wind state. The top row shows the major blocks in the pipeline and the bottom row shows the steps in the SHNO block.}
    \label{fig:figure1}
\end{figure*}

The objective is to propagate the solar wind plasma from a predefined inner boundary at 20 solar radii ($R_\odot$), evolving all eight MHD variables through the heliosphere in a manner consistent with the underlying MHD physics. We achieve this using a spherical harmonic neural operator (SHNO) that learns the radial evolution operator autoregressively, conditioned on the accumulated radial history of the solution. The proposed framework realizes this through a physically motivated learning and a progressive training curriculum, and is outlined in the following sections.

\subsection{Simulation Data and Preprocessing}

The dataset is constructed from 88 steady state 3D solar wind simulations in the inner heliosphere generated with the Alfven Wave Solar Atmosphere Model (AWSoM), spanning a full solar cycle. AWSoM is open source and well validated in many applications \citep{Sachdeva_2021,Szente_2023,Shi_2024,Chen_2025,Sachdeva_2026}. The steady state solution is obtained in the Sun centered inertial frame. The simulations are performed by changing the Poynting flux parameter, which controls the energy input of the model \citep{Jivani:2023,Huang_2023,Huang:2024GONG}, for 2 Carrington rotations per year between 2007 and 2025. They are ordered chronologically and divided into 70 training, 9 validation, and 9 held-out test simulations. 

The native AWSoM simulations provide the eight MHD variables on a 231$\times$181$\times$360 spherical ($r, \theta, \phi$) grid from 20 to 250 $R_\odot$. To reduce the computational cost, the angular grid is downsampled to 91$\times$180, while the model advances the solution using a  fixed radial increment of $\Delta r = 10 R_\odot$. For the analysis presented here, the final prediction at the outer (250$R_\odot$) boundary is excluded to avoid potential boundary related artifacts, leaving 22 reported transitions and 23 shells from 20 to 240 $R_\odot$ (1.12 AU). To increase the diversity of the training samples, each radial sequence is subjected to a random periodic longitudinal shift selected from the 180 unique longitude cells. 

Two successive scalings are applied to prepare the data for learning. First, each variable is nondimensionalized using a variable-specific reference scale, bringing their otherwise widely different physical magnitudes into comparable numerical ranges. A radial rescaling is then applied to compensate for the systematic power-law decay of density, pressure, and magnetic-field strength with heliocentric distance. Each property ($q_i$) is multiplied with $(r/r_0)^{a_i}$, where 

\begin{equation} \label{eq:1}
    r_0 = 20R_{\odot},
\end{equation}
\begin{equation} \label{eq:2}
    q = [\rho, V_r, V_{\theta}, V_{\phi}, p, B_r, B_{\theta}, B_{\phi}],
\end{equation}
\begin{equation} \label{eq:3}
    a = [2, 0, 0, 1, 3, 2, 2, 1].
\end{equation}
    
This prevents the network from spending its capacity on reproducing this known background variation. Density, radial velocity, and pressure are represented in logarithmic form to accommodate their dynamic range and to ensure non-zero values during autoregression. Finally, each transformed variable is z-score normalized using statistics computed from the training set.

\subsection{Model Architecture}

The architecture of SNOW is designed to learn the nonlinear correction to a physically simple approximation of radial solar-wind propagation. SNOW combines three components: rotational ballistic approximation, a spherical harmonic neural operator (SHNO), and causal radial Transformer that summarizes the previously encountered radial history.

As the solar wind propagates outward, its finite radial travel time and the rotation of the Sun produce a systematic longitudinal displacement of plasma structures between neighboring shells. This behavior is constructed through a rotational ballistic approximation ($\mathcal{R}_{\rm BA}$) of the next shell before applying any learned correction. The exact implementation is given in Appendix \ref{app:rba}.

The model then learns the residual correction needed to transform this approximated state into the correct next shell. This learning happens through the SHNO with 64$\times$80 spectral modes. The spherical harmonic basis naturally matches the geometry of the system and provides global angular coupling directly on spherical shells. 

However, the local shell alone does not carry enough information for accurate long range propagation. The errors introduced at one radius can alter the phase, width, and interaction of structures at later radii. We address this with a causal Transformer that summarizes the low-resolution ($16\times32$ angular grid) radial history already encountered by the model and supplies this context ($\mathcal{C}_k$) to the SHNO. During full autoregressive rollout, this history is built from previously predicted states.

For a given ground truth shell ($\mathbf{S}_k$) at a radial distance of $r_k$, the next prediction shell ($\widehat{\mathbf{S}}_{k+1}$) will be:

\begin{equation}
    \widehat{\mathbf{S}}_{k+1}
    =
    \mathbf S^{\rm FA}_{k+1} + \Delta \mathbf S^{\mathrm{SHNO}_{k+1}}
\end{equation}

where,

\begin{equation}
    \mathbf S^{\rm FA}_{k+1} = \mathcal{R}_{\rm BA}(\mathbf{S}_k, \Delta r) ,
\end{equation}

\begin{equation}
    \Delta{\mathbf S}^{\mathrm{SHNO}_{k+1}} = \mathrm{SHNO}\!
    \left(
    \mathbf S^{\rm FA}_{k+1}, \mathbf{C}_k, r_k, \Delta r
    \right).
\end{equation}

The first approximation of next shell ($\mathbf S^{\rm FA}_{k+1}$) captures simple rotational transport but not the non-linear MHD effects like radial acceleration, nonlinear stream interaction, or magnetic field evolution. The SHNO learns these missing effects as a residual correction ($\Delta \mathbf S^{SHNO}_{k+1}$). Its input contains 58 channels: 8 transported MHD variables, 48 radial context channels ($\mathbf{C}_k$), current radius ($r_k$), and radial step size ($\Delta r$).

\subsection{Training Curriculum}

A model trained only to minimize one-step error can still become unstable once its own predictions are fed back as inputs. This is especially important here because repeated autoregressive shell to shell transitions are required across the domain. Training SNOW therefore has two simultaneous goals: reproducing the next AWSoM shell accurately, and remaining stable as prediction errors accumulate over radial distance.

The training configuration schedules 420 epochs (E) using a progressive rollout curriculum. The active horizon ($K$), defined as the number of consecutive shell transitions included in a training sequence, is progressively increased during training so that the model learns stable autoregressive propagation. Thus, $K=1$ trains a single $10R_\odot$ transition, whereas $K=22$ spans the complete $20-240R_\odot$ domain.

The whole curriculum can be understood as five conceptual phases across the epochs. SNOW first learns the local residual shell to shell mapping without radial context (\textit{residual} phase: 0-49). The causal Transformer is then introduced to provide information from the inner-shell history (\textit{context} phase: 50-119). Next, the model learns to correct states that have drifted away from the AWSoM trajectory (\textit{recovery} phase: 120-164). Gradients are subsequently propagated through short chains of successive predictions (\textit{connected rollout} phase: 165-209). Finally, the rotational approximation is included in this connected gradient path while the horizon increases to the complete domain (\textit{differentiable rotation} phase: 210-419). Once introduced, each mechanism remains active during subsequent stages.

The primary training loss is a channel-weighted Huber loss \citep{huber_robust_1964} evaluated using ground-truth shells and histories. This term anchors the local shell to shell evolution throughout training. To expose SNOW to the longer rollout errors, two additional loss terms are activated at later curriculum phases, one associated with \textit{recovery} and the other with \textit{connected rollout}. The former teaches the model how to recover from accumulated drift, whereas the latter directly penalizes the compounding of errors across multiple radial steps. Together, these objectives allow the local evolution operator to remain accurate while improving stability under full autoregressive propagation.

Online validation is performed in two complementary ways. During the curriculum, performance is evaluated over the active rollout horizon, K(E). In parallel, fixed full-march validation is performed from the inner boundary across the complete radial domain. This forecast is initialized only with the 20 $R_\odot$ AWSoM shell and uses model-generated states at every subsequent radius. This provides a common measure of long-range performance as the curriculum changes. After the architecture and checkpoint are fixed, the same full-march protocol is applied once to the nine previously unseen test simulations.

SNOW contains approximately \(128.3\) million trainable parameters. Both the SHNO and the Transformer components are optimized jointly using AdamW \citep{Loshchilov_adamw}, with batch size eight per GPU, initial learning rate \(3\times10^{-4}\), and weight decay \(10^{-4}\).

\section{Results} \label{sec:results}
We evaluate SNOW on nine held-out AWSoM simulations that were not used during training, validation, or model selection. Each test case is initialized only with the $20R_\odot$ inner-boundary state, after which all shells out to $240R_\odot$ are generated autoregressively without access to intermediate AWSoM states. We assess the resulting solutions in terms of global 3D reconstruction, vector orientation and magnetic topology, and preservation of bulk-flow structures across the complete rollout. All the results were generated using the best checkpoint, which was after completion of 400 epochs.

\subsection{3D Solution with Rollout}

\begin{figure*}
    \centering
    \includegraphics[width=\textwidth]{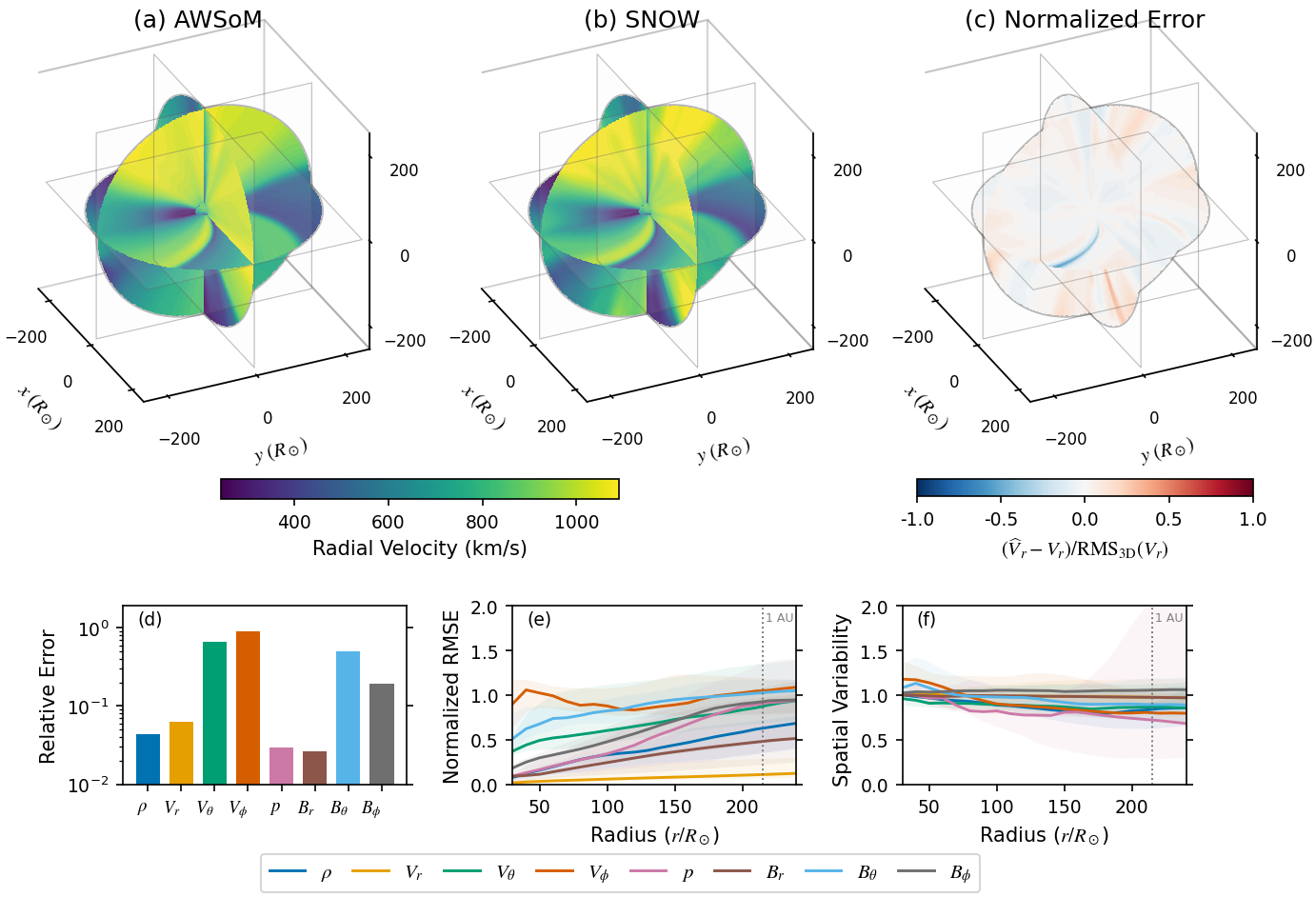}
    \caption{Global 3D reconstruction and rollout performance of SNOW. Subplots (a–c) illustrate the AWSoM, SNOW, and their residual for a representative held-out realization. Panels (d–f) show the full-domain reconstruction error, radial error accumulation, and spatial-variability ratio. Curves in (e–f) summarize the nine held-out test simulations.
    }
    \label{fig:figure2}
\end{figure*}

Figure 2 presents the 3D radial velocity field after the complete autoregressive rollout for both AWSoM and SNOW. The global view of the heliospheric structure is visualized using three orthogonal slices, one equatorial plane and two mutually perpendicular meridional planes. To visualize the local prediction error, Figure 2(c) shows the normalized error at each grid point:

\begin{equation}
    E_{V_r}({r,\theta,\phi}) =
    \frac{\widehat{V}_r({r,\theta,\phi})-V_r({r,\theta,\phi})}
    {\mathrm{RMS}_{3\mathrm{D}}(V_r)},
    \label{eq:5}
\end{equation}
where $\mathrm{RMS}_{3\mathrm{D}}(V_r)$ is the global root mean square AWSoM radial velocity evaluated over all grid points in the complete 3D domain.

SNOW reproduces the principal fast and slow wind regions, including their longitudinal curvature, angular extent, and radial evolution. The largest normalized residuals are concentrated along stream boundaries, while most of the domain exhibits substantially smaller errors. Paired positive and negative residuals along these interfaces are consistent with modest displacements or changes in structure width rather than the disappearance of the underlying large scale features.

Figure 2(d) summarizes the 3D reconstruction accuracy of all eight MHD variables using the relative \(L_2\) error:

\begin{equation}
    E_c^{\mathrm{3D}} = \left[ \frac{ \sum_{i,j,\ell} \left[ \widehat{S}_c(r_i,\theta_j,\phi_\ell) - S_c(r_i,\theta_j,\phi_\ell) \right]^2 }{ \sum_{i,j,\ell} S_c^2(r_i,\theta_j,\phi_\ell) } \right]^{1/2}
\end{equation}

Density, radial velocity, thermal pressure, and radial magnetic field show the lowest errors, each remaining below approximately \(0.1\) for the representative realization. Larger relative errors occur for the transverse velocity and magnetic field components, particularly $V_\theta$, $V_\phi$, and $B_\theta$.

Radial accumulation of prediction error is quantified in Figure 2(e) using the relative $L_2$ error. The normalization is evaluated independently on each spherical shell. The curves show the median across the nine held-out simulations, with the shaded regions indicating the corresponding test-set spread. The errors generally increase with heliocentric distance as predictions are repeatedly supplied back to the model. This increase remains gradual across the complete rollout, with no abrupt increase indicating loss of autoregressive stability.

Finally, we evaluate whether the model preserves the amplitude of spatial variations rather than converging toward an overly smooth solution. Figure 2(f) shows the spatial-variability ratio ($A_c(r)$),

\begin{equation}
    A_c(r_k) = 
    \frac{
    \sigma_{\theta,\phi}
    \left[
    \widehat{S}_c(r_k,\theta,\phi)
    \right]
    }{
    \sigma_{\theta,\phi}
    \left[
    S_c(r_k,\theta,\phi)
    \right]
    },
\end{equation}
where $\sigma_{\theta,\phi}$ is the standard deviation on a spherical shell at $r_k$. A value of unity indicates that SNOW and AWSoM contain the same amplitude of spatial variability, while values below or above unity indicate reduced or enhanced variability, respectively. The median ratios remain broadly close to unity through most of the domain, with variable-dependent deviations toward larger heliocentric distances. Pressure exhibits largest case-to-case spread in the outer heliosphere, with its 1$\sigma$ equivalent interval spanning 0.30-2.09.

\begin{figure*}
    \centering
    \includegraphics[width=\textwidth]{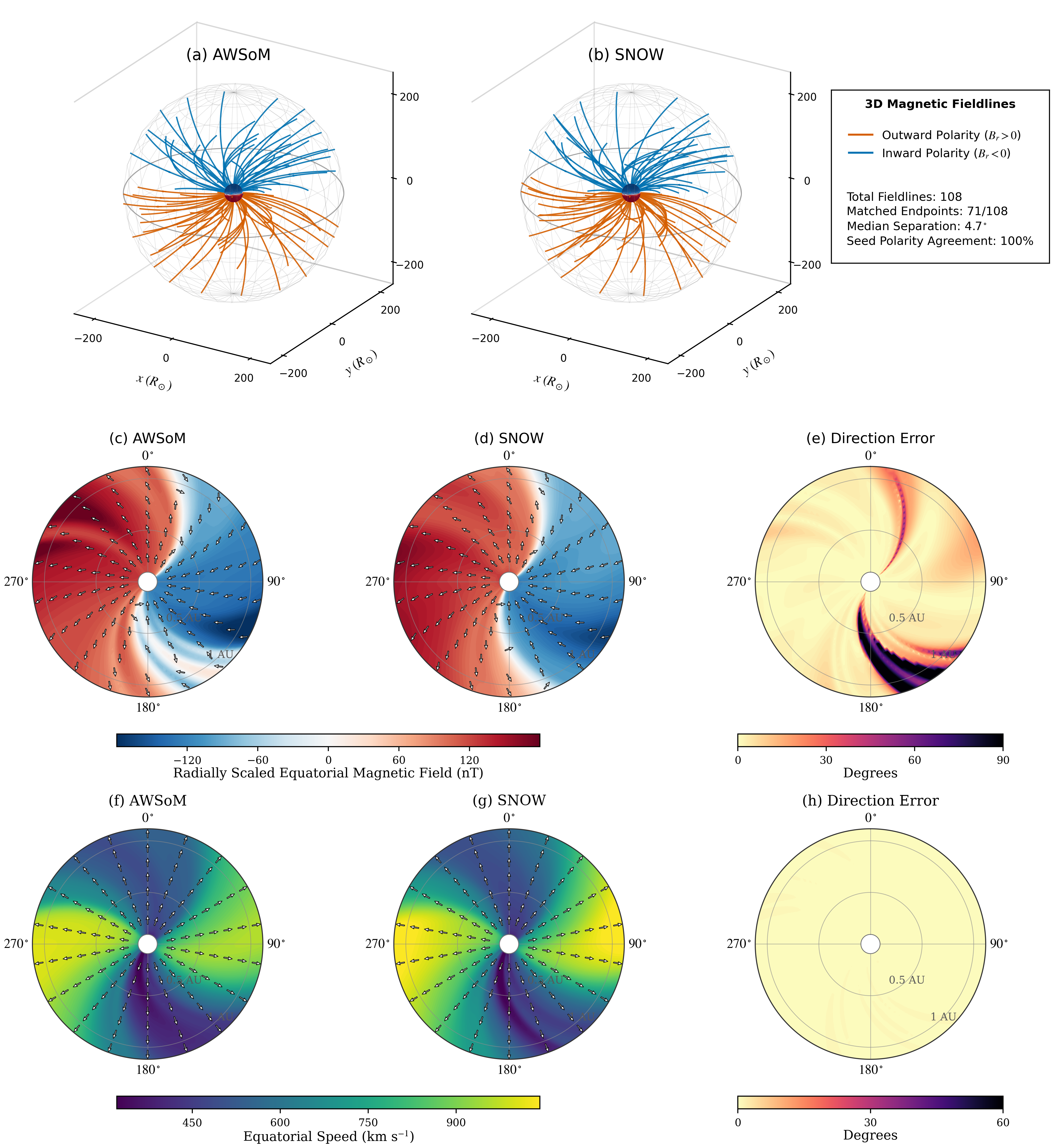}
    \caption{Magnetic topology and projected vector orientation reconstructed by SNOW. Panels (a-b) represent 3D magnetic fieldlines traced from identical seed locations in AWSoM and SNOW. Subplots (c-e) are projected magnetic field vectors and angular error, and (f-h) show the corresponding velocity comparison.
    }
    \label{fig:figure3}
\end{figure*}

\begin{figure*}
    \centering
    \includegraphics[width=\textwidth]{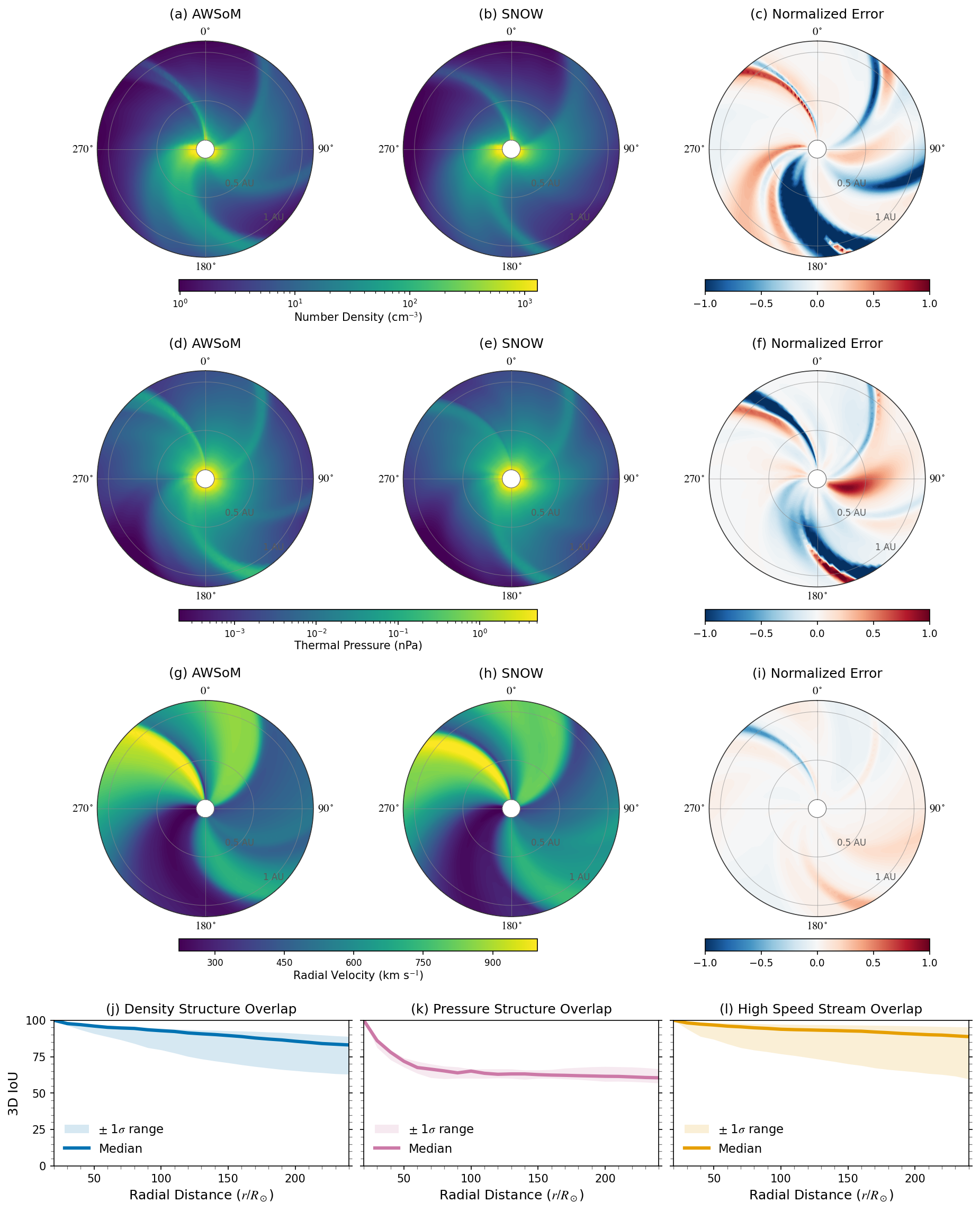}
    \caption{Preservation of bulk solar-wind structures during the autoregressive rollout. Panels (a–i) show the equatorial AWSoM, SNOW, and residual maps for proton density, thermal pressure, and radial velocity. Subplots (j–l) are shell-wise IoU of the corresponding structures across the nine test simulations.}
    \label{fig:figure4}
\end{figure*}

\subsection{Vectors and Field Lines}
Figure 3(a-b) compares 3D magnetic field lines traced from identical seed locations at $20R_\odot$ in the representative test realization. SNOW accurately reproduces the large scale inward and outward polarity sectors and their winding through the heliosphere. Of the 108 seeded field lines, 71 produce matched outer-boundary endpoints in both solutions. For these matched trajectories, the median angular endpoint separation is 4.7$^\circ$, while the magnetic polarity at the seed locations agrees in 100\% of cases. These diagnostics quantify the agreement in the magnetic connectivity represented by the two 3D solutions.

Figures 3(c-e) compare the magnetic field vectors projected onto the equatorial \((r,\phi)\) plane. For visualization, the radial magnetic field ($B_r$) is scaled by $(r/r_0)^2$ to remove its dominant radial decay.

The directional difference ($\Delta\psi_B$) between the projected AWSoM and SNOW magnetic vectors is

\begin{equation}
    \Delta\psi_B
    =
    \cos^{-1}
    \left[
    \frac{
    B_r\widehat{B}_r+B_\phi\widehat{B}_\phi
    }{
    \sqrt{B_r^2+B_\phi^2}
    \sqrt{\widehat{B}_r^2+\widehat{B}_\phi^2}
    }
    \right],
\end{equation}

reported in degrees. SNOW reproduces the broad positive and negative $B_r$ sectors and their associated projected field directions. The largest angular differences occur primarily in narrow regions surrounding polarity reversals and strongly curved sector boundaries, while substantially smaller errors occur throughout most of the remaining equatorial domain.

Figures 3(f-h) show the corresponding comparison for the projected plasma velocity in $r-\phi$ plane. The same angular metric is used to quantify the difference between the projected velocity vectors. The AWSoM and SNOW velocity directions agree closely across most radii and longitudes, with localized increases in directional error near stream interfaces.

\subsection{Bulk Flow Properties}

Figure 4 compares equatorial slices of proton number density, thermal pressure, and radial velocity for a second representative held-out realization. SNOW reproduces the large scale spiral morphology of the enhanced density and pressure regions together with the corresponding fast and slow wind streams. 

The same normalized residual form as Equation \ref{eq:5} is used, but the RMS normalization is evaluated independently at each radius. For density and pressure, the largest residuals are concentrated primarily along narrow compression structures and stream interfaces, while the radial-velocity field shows comparatively small residuals over most of the equatorial domain.

To quantify the structural agreement across all nine held-out cases, Figures 4(j–l) show a shell-wise intersection over union (IoU). For binary AWSoM and SNOW structure masks $\widehat{\mathcal{M}}$ and corresponding AWSoM structure $\mathcal{M}$,

\begin{equation}
    IoU(r_k) = 
    100 \frac{\sum_{j\ell}
    \left[
    \widehat{\mathcal{M}}_{j\ell}(r_k)
    \cap
    \mathcal{M}_{j\ell}(r_k)
    \right]
    }{\sum_{j\ell}
    \left[
    \widehat{\mathcal{M}}_{j\ell}(r_k)
    \cup
    \mathcal{M}_{j\ell}(r_k)
    \right]
    }.
\end{equation}

Thus, 100\% indicates identical angular structure on a given spherical shell. Density and pressure structures are defined as regions exceeding the 50th percentile of AWSoM value on 3D solution, whereas high speed streams are defined by $V_r>600~\mathrm{km,s^{-1}}$. The curves show the median IoU across all nine held-out simulations, and the shaded regions indicate the $16^\mathrm{th}-84^\mathrm{th}$ percentile interval.

The density structure overlap decreases gradually with radius and remains approximately 83\% in the outer heliosphere. The pressure overlap decreases more rapidly over the inner part of the rollout before approaching approximately 60\%, whereas the high speed stream overlap remains close to 90\% across most of the radial domain. Thus, among the three structures considered here, SNOW shows the highest structural overlap for high speed streams, followed by density enhancements and pressure structures.

\section{Discussion} \label{sec:discussion}

\begin{figure*}
    \centering
    \includegraphics[width=\textwidth]{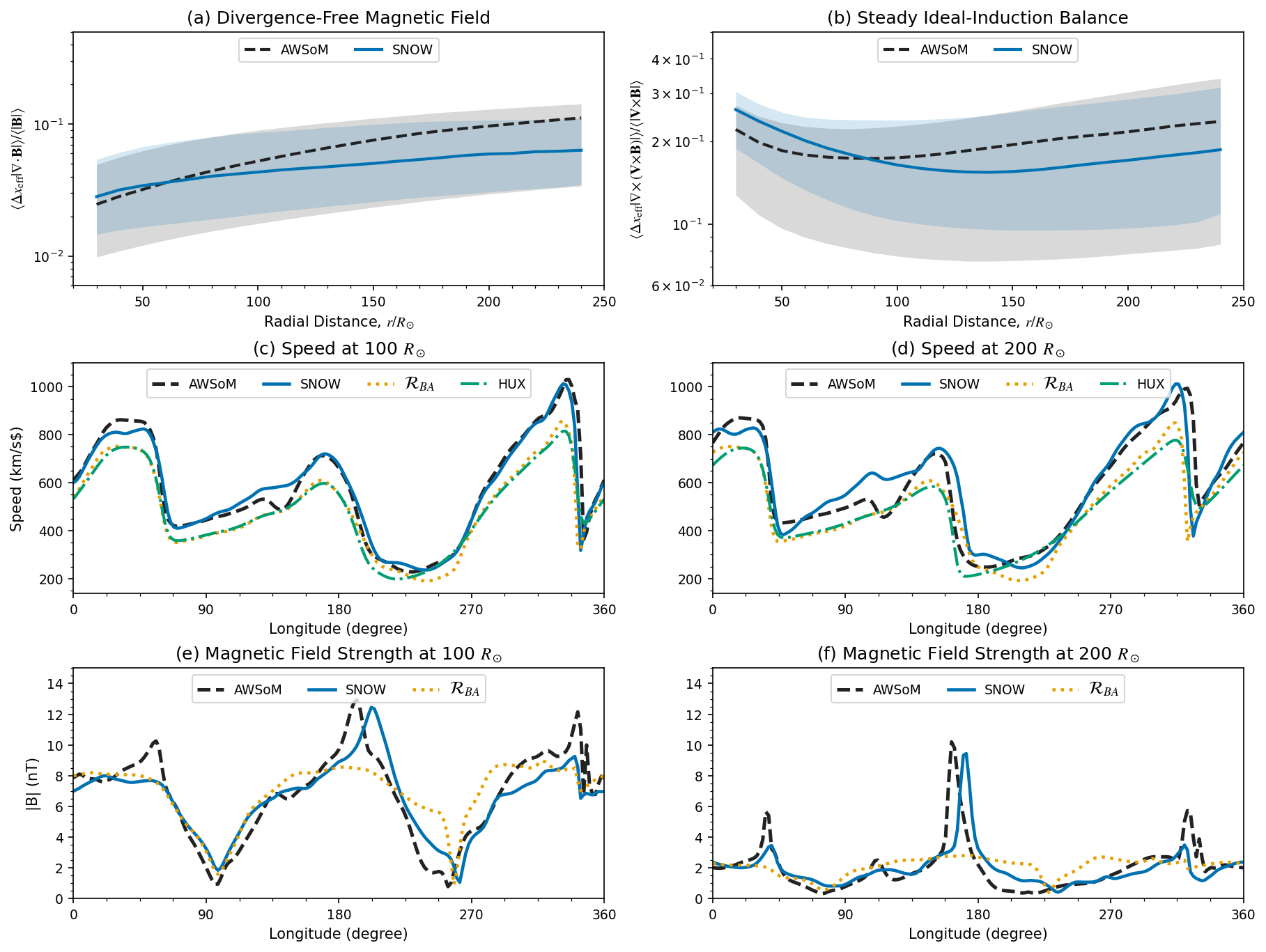}
    \caption{Physical diagnostics and comparison with reduced propagation models. Top row: (a–b) shows the magnetic divergence and induction balance diagnostics for AWSoM and SNOW. Panels (c–d) are radial velocity profiles at \(100\) and \(200\,R_\odot\) compared with the rotational ballistic first approximation ($\mathcal{R}_{BA}$) and HUX. Panels (e–f) are corresponding magnetic field strength profiles. 
    }
    \label{fig:figure5}
\end{figure*}

The results in Section \ref{sec:results} demonstrate that SNOW can reconstruct the global 3D steady state solar wind consecutive autoregressive radial steps while retaining the major scalar, vector, and structural characteristics of the reference MHD solutions. The importance of this result extends beyond pointwise agreement with AWSoM. A useful heliospheric surrogate must remain stable over long propagation distances, preserve the physical organization of the solar wind, and provide outputs that can be incorporated into broader forecasting workflows. We therefore discuss the physical fidelity of the learned evolution, its potential role within an operational modeling context, and the present limitations and future extensions of the framework.

\subsection{Physical Fidelity}

A central challenge in autoregressive surrogate modeling is the accumulation of prediction errors. Figure \ref{fig:figure2}(e) shows that the error increases gradually with heliocentric distance rather than exhibiting an abrupt divergence at a particular radius. This behavior is important because every predicted shell becomes the input to the subsequent radial step, allowing local errors to propagate through the remainder of the solution. The smooth error growth over the complete 22-step rollout therefore indicates that SNOW maintains stable radial propagation despite repeatedly operating on its own predictions.

This stability is accompanied by preservation of the dominant physical organization of the heliosphere. Figures 2–4 show that SNOW retains the fast and slow wind streams, angular variability, magnetic polarity sectors, sampled magnetic connectivity, projected vector orientations, and bulk flow structures of the AWSoM solutions. In particular, the spatial variability ratio remains broadly close to unity, showing that stable propagation is not achieved simply by progressively smoothing the solution. Similarly, the recovery of magnetic-sector structure and fieldline connectivity indicates that agreement among the individual MHD variables translates into coherent vector field organization.

Independent physical diagnostics in Figure 5 provide an additional test of this interpretation. The normalized magnetic divergence measure obtained from SNOW remains of the same order as AWSoM throughout the modeled domain, with substantial overlap between the two distributions. The induction balance diagnostic similarly follows the broad radial behavior of the reference solution, including its evolution through the inner and outer heliosphere. Neither diagnostic is explicitly constrained during training. Their comparable magnitude and radial trends therefore provide independent evidence that the learned rollout retains important magnetic field couplings present in the underlying MHD solutions. These comparisons do not imply exact satisfaction of the MHD equations or constitute a formal conservation guarantee, but they strengthen the physical credibility inferred from the structural diagnostics in Figures 2–4.

\subsection{Operational Context}

From a concept-of-operations perspective, SNOW occupies the heliospheric propagation domain of a space weather modeling chain. Its input is a complete MHD state at the inner boundary, $20R_\odot$, from which it generates the 3D evolution of all eight plasma properties out to $240R_\odot$. A complete rollout requires approximately $<$1 s on a single GPU, compared with $>$100 CPU hours for the corresponding global MHD solving models like AWSoM and ENLIL. An upstream coronal or inner-boundary model can therefore provide the initial plasma state, SNOW can perform the heliospheric propagation, and the resulting plasma and magnetic field conditions can subsequently be extracted at Earth or other locations for use by downstream forecasting models. The reduction in computational cost also makes ensemble-based calculations substantially more practical.

Figures \ref{fig:figure5}(c–d) provide a direct comparison with reduced solar wind propagation approaches. At $100R_\odot$, SNOW reconstructs the AWSoM radial velocity profile with an RMSE of 58.3 $\mathrm{km/s}$, compared with 101.9 $\mathrm{km/s}$ for the rotational first approximation and 99.1 $\mathrm{km/s}$ for HUX. At $200R_\odot$, the corresponding errors are 83.8, 111.8, and 102.3 $\mathrm{km/s}$, respectively. Additionally, figures \ref{fig:figure5}(e–f) show a similar comparison for magnetic field strength. SNOW therefore reproduces the MHD velocity and magnetic structures more accurately than these simpler propagation schemes at both intermediate and near-Earth heliocentric distances, while retaining a similarly lightweight computational character. 

More importantly, reduced models such as HUX propagate primarily the solar wind speed, whereas SNOW simultaneously evolves the complete 3D plasma and magnetic field state. The same calculation can therefore provide density, pressure, the full velocity vector, and magnetic field components required by subsequent geospace models rather than only a speed boundary condition.

A further distinction is that the complete SNOW propagation is differentiable. Gradients can therefore, in principle, be propagated from an output location through the heliospheric evolution toward the inner-boundary state or other model inputs. This creates a route toward gradient-based data assimilation, sensitivity analysis, inverse estimation, and optimization within coupled forecasting systems. These capabilities are not exercised in the present study and should therefore be regarded as enabled applications rather than demonstrated results. In this role, SNOW exemplifies the possible evolution of operational space weather forecasting away from large numerical simulation models and towards extremely fast surrogate models. 

\subsection{Limitations and Future Directions}
Although SNOW reproduces the large scale heliospheric solution with good overall fidelity, the reconstruction accuracy is not uniform across all physical regimes. The largest residuals are generally concentrated around stream interfaces and compressed structures, where relatively small positional or amplitude differences produce larger local errors. In addition, the autoregressive formulation inevitably allows small prediction errors to accumulate with radial distance. The present curriculum prevents this accumulation from developing into an instability, but the gradual degradation toward the outer heliosphere remains an important target for further improvement. Beyond these performance characteristics, the present framework does not incorporate the temporal evolution and therefore cannot evolve transients such as CME propagation. Finally, although the 88 simulation ensemble is sufficient to demonstrate the feasibility of the approach, a substantially larger and more diverse ensemble is required for a systematic assessment of generalization beyond the present test distribution.

These limitations also define the immediate directions for extending the framework. A natural extension is to introduce temporal autoregression in addition to radial propagation, transforming the present steady state surrogate into a fully dynamic 3D MHD surrogate. Such a formulation would enable the evolution of CMEs. 

Training on a larger and more diverse ensemble of MHD simulations will enable a more systematic assessment of generalization and robustness. A necessary next step is to compare SNOW, HUX, ENLIL, and AWSoM-IH for common events and consistent inner-boundary conditions against in-situ observations. Such a comparison will distinguish the surrogate error relative to AWSoM from the model-to-observation uncertainty of the complete forecasting chain. SNOW can also be coupled with WSA+ to form a differentiable
Sun-to-Earth heliospheric pipeline.

The present study focuses on global physical fidelity and benchmarking the model's operational capability. Detailed architectural, optimization, and curriculum ablations will be published as a separate methodological study.

\section{Conclusion} \label{sec:conclusion}
In this work, we present SNOW, a fast and differentiable surrogate for global 3D steady state MHD solar wind evolution. Starting from a single inner-boundary state at $20R_\odot$, SNOW autoregressively evolves all eight MHD variables through the heliosphere to $240R_\odot$, while maintaining stable long-range propagation and preserving the large scale plasma and magnetic field organization of the reference solutions. The results demonstrate that the complete 3D MHD state of the heliosphere can be emulated with physically meaningful fidelity at a computational cost of less than a second. This establishes full-state neural emulation as a viable approach for accelerating global heliospheric modeling while retaining the physical information provided by conventional MHD simulations.

\begin{acknowledgments}
This research was supported by the NASA Living with a Star Jack Eddy Postdoctoral Fellowship Program, administered by UCAR’s Cooperative Programs for the Advancement of Earth System Science (CPAESS) under award \#80NSSC22M0097.  This work was partially supported by NASA under awards No 80NSSC23M0192, 80NSSC20K1580, 80NSSC21K1555. AI tools were used to assist with the refinement of the manuscript text and code used in this study.
\end{acknowledgments}

\textit{Code Sharing Statement:} 
All codes developed in this paper, along with the SNOW model checkpoint, are archived at https://doi.org/10.5281/zenodo.23090129.

\software{PyTorch \citep{NEURIPS2019_9015}.}

%% Appendix material should be preceded with a single \appendix command.
%% There should be a \section command for each appendix. Mark appendix
%% subsections with the same markup you use in the main body of the paper.
%%
%% Each Appendix (indicated with \section) will be lettered A, B, C, etc.
%% The equation counter will reset when it encounters the \appendix
%% command and will number appendix equations (A1), (A2), etc. The
%% Figure and Table counter will not reset.

\appendix

\section{Rotational Ballistic Approximation}
\label{app:rba}

SNOW learns each radial shell as a correction to a rotational ballistic
first approximation, $\mathcal{R}_{\rm BA}$. This baseline consists of radial
extrapolation and longitudinal rotation of the solar-wind state across
$20$--$240\,R_\odot$. For the radial extrapolation, the state variables are
assumed to follow the channel-dependent radial power laws described in
Equations~(\ref{eq:2}) and~(\ref{eq:3}).

For the longitudinal rotation, the displacement is calculated from the radial
velocity. At each latitude, its longitudinal mean, $\overline V_r$, and a
centered five-point periodic moving average, $\widetilde V_r$, are calculated:

\begin{equation}
\overline V_r(\theta)
=
\frac{1}{N_\phi}
\sum_j V_r(\theta,\phi_j),
\end{equation}

\begin{equation}
\widetilde V_r(\theta,\phi_j)
=
\frac{1}{5}
\sum_{m=-2}^{2}
V_r\left(\theta,\phi_{(j+m)\bmod N_\phi}\right).
\end{equation}

The corresponding longitudinal displacement is constructed as

\begin{align}
\Delta\phi_{\rm BA}(\theta,\phi)
={}&
\frac{\Omega_\odot\Delta r}{\overline V_r(\theta)}
+
\frac{\Omega_\odot\Delta r}{2}
\left[
\frac{1}{\widetilde V_r(\theta,\phi)}
-
\frac{1}{\overline V_r(\theta)}
\right].
\label{eq:rba_shift}
\end{align}

The averaging suppresses the discontinuities and extreme displacements
produced by a pointwise $1/V_r$ mapping while retaining the resolved
fast--slow stream structure. Because the SHNO correction is initially zero,
training begins from this smooth and kinematically meaningful baseline. This
is especially useful during the early rollout stages and limits the
accumulation of numerical artifacts as the rollout length increases through
the training curriculum.

\bibliography{SNOW}{}
\bibliographystyle{aasjournalv7.1}

%% This command is needed to show the entire author+affiliation list when
%% the collaboration and author truncation commands are used.  It has to
%% go at the end of the manuscript.
%\allauthors

%% Include this line if you are using the \added, \replaced, \deleted
%% commands to see a summary list of all changes at the end of the article.
%\listofchanges

\end{document}